\documentclass[prd,twocolumn,superscriptaddress,showpacs,nofootinbib,preprintnumbers]{revtex4}

\usepackage{amsmath}
\usepackage{amsfonts}
\usepackage{graphicx}
\usepackage{dcolumn}
\usepackage{hyperref}
\def\bea{\begin{eqnarray}}
\def\eea{\end{eqnarray}}
\def\beqa{\begin{equation}}
\def\eeqa{\end{equation}}

\def\be{\begin{equation}}
\def\ee{\end{equation}}

\def\5{\overline 5}
\def\vp{\varphi}

\begin{document}

\title{
What is needed of a  tachyon if it is to be the dark energy?  }
\author{Edmund J.~Copeland}
\affiliation{Department of Physics and Astronomy,
University of Sussex, Falmer, Brighton BN1 9QJ, United Kingdom}
\author{Mohammad R.~Garousi}
\affiliation{Department of Physics, Ferdowsi University, P.O.Box 1436,
Mashhad, Iran\\ and\\
Institute for Studies in Theoretical Physics and Mathematics IPM\\
P.O.Box 19395-5531, Tehran, Iran}
\author{M.~Sami}
\affiliation{IUCAA, Post Bag 4, Ganeshkhind,
Pune 411 007, India}
\author{Shinji Tsujikawa}
\affiliation{Department of Physics, Gunma National College of
Technology, Gunma 371-8530, Japan}

\date{\today}

\vskip 1pc
\begin{abstract}
We study a dark energy scenario in the presence of a tachyon
field $\phi$ with potential $V(\phi)$ and
a barotropic perfect fluid.
The cosmological dynamics crucially depends on the
asymptotic behavior of the quantity
$\lambda=-M_pV_\phi/V^{3/2}$.
If $\lambda$ is a constant, which corresponds to an
inverse square potential $V(\phi) \propto \phi^{-2}$,
there exists one stable critical point that gives an
acceleration of the universe at late times.
When $\lambda \to 0$ asymptotically, we can have
a viable dark energy scenario in which the system
approaches an ``instantaneous'' critical point that
dynamically changes with $\lambda$.
If $|\lambda|$ approaches infinity asymptotically,
the universe does not exhibit an acceleration at late
times. In this case, however, we find an interesting
possibility that a transient acceleration occurs
in a regime where $|\lambda|$ is smaller than
of order unity.
\end{abstract}

\pacs{98.70.Vc, 98.80.Cq}

\maketitle
\vskip 1pc

\section{Introduction}

There are few more taxing questions facing cosmology today 
than what is the nature of the dark energy in the Universe? 
This almost uniform distribution of energy density with a 
substantial negative pressure completely dominates all other 
forms of matter and yet the best we can do is to infer its existence
from data (see Refs.~\cite{review} for reviews). 
Moreover, because today it appears to behave just 
like a cosmological constant, we are naturally led to take 
seriously the possibility  that what we are observing is a remnant 
of the physics of the very early Universe! 
 
Over the past few years there have been many papers devoted 
to addressing the nature of the dark energy. Explanations 
include: a true cosmological constant possibly arising from 
a Landscape type picture of string vacua \cite{KKLT};  
dynamical `Quintessence' fields \cite{Paul} which have tracker like 
properties where the energy density in the fields track those 
of the background energy density before dominating 
today; K-essence scenarios \cite{Kes}, 
where the acceleration is driven by modified kinetic terms 
in the underlying action; modifications of gravity \cite{branquint}
(mainly motivated by Braneworld models) which
lead to late time accelerating solutions of the modified Friedmann 
equation and Chaplygin gas models \cite{gas} 
which attempt to incorporate a unified description of  dark energy
and dark matter. The list goes on, but the aim of all the models is the same,
to explain why the Universe has only recently started accelerating 
with an energy density so close to the critical density.

In this paper, we turn our attention to the issue of the tachyon
as a source of the dark energy. The tachyon is an unstable field 
which has become important in string theory through its role in the
Dirac-Born-Infeld (DBI) action which is used to describe the
D-brane action \cite{as1,s1}. A number of authors have already
demonstrated that the tachyon could play a useful role in
cosmology \cite{Tachyonindustry}, 
independent of the fact that it is an unstable field. 
It can act as a source of dark matter and
can lead to a period of inflation depending on the form of the
associated potential. Indeed it has been proposed as the source
of dark energy for a particular class of 
potentials \cite{Paddy,Bagla,AF,AL,GZ}. However,
there has not really been an effort to understand the general
properties of tachyonic cosmologies. We attempt to do that here.
Starting with the four-dimensional Dirac-Born-Infeld (DBI) 
action, the tachyon field (with arbitrary potential
$V(\phi)$) is coupled to a background perfect fluid of radiation
or matter. Modifying the procedure introduced in \cite{CLW} (see
also 
Refs.\,\cite{vandenHoogen:1999qq,Billyard:1998hv,
LS,MP,NNR,MM,Maeda01,Huey:2001ae,Mizuno,PT,SST} 
on a related theme), the evolution equations for the system of tachyon plus
a background fluid can be written as first order differential
equations involving two variables $x$ and $y$ where $x \propto
\dot{\phi}$ and $y \propto \sqrt{V(\phi)}/H$, $H$ being the Hubble
parameter. Such equations have solutions whose behaviour
depends crucially on the quantity
$\lambda=-M_pV_\phi/V^{3/2}$, where $V_\phi \equiv 
{\rm d}V/{\rm d}\phi$. We
analyse this behaviour for a wide class of potentials. Depending
on the asymptotic behaviour of $\lambda$ we either obtain late
time attractor solutions which correspond to the well known
inflationary cosmologies associated with the tachyon potential
$V(\phi) \propto 1/\phi^2$, or perhaps more interesting
cosmologies which include (for $\lambda \to 0$) asymptotic
behaviour where the system approaches an ``instantaneous''
critical point that dynamically changes with $\lambda$. For the
case where $|\lambda| \to \infty$ asymptotically, the system does
not lead to late time acceleration. However, it does show a
period of transient acceleration in the regime where $|\lambda| <
{\cal O}(1) $. It raises the possibility that we are living in such a
transient regime.

The rest of the paper is as follows: Section II introduces the
DBI action and the associated equations of motion for the tachyon
field including the background perfect fluid. The equations of
motion in terms of the new variables $x$, $y$ and $\lambda$ are
presented along with the effective equation of state for the
tachyon. Potentials which demonstrate the particular asymptotic
features of $\lambda = {\rm const}, \, 0$ or $\infty$ are
included. Section III considers the particular case of constant
$\lambda$ finding all the associated fixed points for the
equations of motion and demonstrating their stability before going
on to show how the existence restricts the allowed form of the
potential for the tachyon. We show how easy it is to obtain late
time accelerating solutions in this case. Section IV considers
the case $\lambda \to 0$ for both a non-oscillating and
oscillating late time evolution of $\phi$, considering a class of
potentials that have these features. As before we find it is
possible to have late time acceleration and classify the
conditions that have to be met by the tachyon potential. Section
V does a similar thing but for a class of potentials which lead
asymptotically to $\lambda \to \infty$. Although generally we do
not have late time inflation, we show there is a novel feature
present which is a transient period of inflation for exponential
potentials. We summarise in Section VI.

\section{DBI Model}

The Dirac-Born-Infeld type effective 4-dimensional action
for our system is described by \cite{as1}
\begin{eqnarray}
\label{action}
{\cal S} &=& \int d^4x\biggl\{\sqrt{-g}
\frac{M_p^2}{2}R \nonumber \\
& & -V(\phi)\sqrt{-\det(g_{\mu\nu}+
\partial_{\mu}\phi\partial_{\nu}\phi)}\biggr\}\,,
\end{eqnarray}
where $M_p$ is the reduced Planck mass, $R$ is the
scalar curvature and $V(\phi)$ is the potential 
of the tachyon field $\phi$.
The above tachyon DBI action is believed to describe the 
physics of tachyon condensation for all values of $\phi$ 
as long as string coupling and the second derivative 
of $\phi$ are small.

We shall consider a cosmological scenario in which
the system is filled with the field $\phi$ and
a barotropic perfect fluid with an equation of state
$p_B=(\gamma-1)\rho_B$.
Note that $\gamma=1$ for a pressureless dust and
$\gamma=4/3$ for radiation.
In a spatially flat Friedmann-Lemaitre-Robertson-Walker (FLRW) 
metric with a scale factor $a$, the pressure and 
the energy densities of the field
$\phi$ are given, respectively, by
\begin{eqnarray}
\label{prho}
p_\phi=-V(\phi)\sqrt{1-\dot{\phi}^2}\,,~~~
\rho_\phi=\frac{V(\phi)}{\sqrt{1-\dot{\phi}^2}}\,.
\end{eqnarray}
The background equations of motion are
\begin{eqnarray}
\label{Hubble} & & \dot{H} =
-\frac{\dot{\phi}^2V(\phi)}{2M_p^2\sqrt{1-\dot{\phi}^2}}
-\frac{\gamma}{2}\frac{\rho_B}{M_p^2}\,,\\
\label{ddphi}
& & \frac{\ddot{\phi}}{1-\dot{\phi}^2}+3H\dot{\phi} +
\frac{V_\phi}{V}=0\,,\\
& & \dot{\rho}_B+3\gamma H \rho_B=0\,,
\label{phi}
\end{eqnarray}
together with a constraint equation for the Hubble 
parameter:
\bea
\label{Hubbleeq}
3M_p^2H^2 =
\frac{V(\phi)}{\sqrt{1-\dot{\phi}^2}}+\rho_B\,.
\eea

Let us rewrite the above equations in an autonomous form.
We define the following dimensionless quantities:
\begin{eqnarray}
\label{Dquantity}
x=\dot{\phi}=H\phi'\,,~~~y=\frac{\sqrt{V(\phi)}}
{\sqrt{3}HM_p}\,,
\end{eqnarray}
where a prime denotes the derivative with respect to
the number of $e$-folds, $N={\rm ln}\,a$.
Then we obtain the following equations:
\begin{eqnarray}
\label{dotx}
& & x'=-(1-x^2)(3x-\sqrt{3}\lambda y)\,, \\
\label{doty}
& & y'=\frac{y}{2}\left(-\sqrt{3}
\lambda xy-\frac{3(\gamma -x^2)y^2}
{\sqrt{1-x^2}}+3\gamma \right) \,,\\
& & \lambda'=-\sqrt{3}\lambda^2 xy(\Gamma-3/2)\,,
\label{auto}
\end{eqnarray}
where
\begin{eqnarray}
\label{lam}
\lambda=-\frac{M_pV_\phi}{V^{3/2}}\,,~~~
\Gamma=\frac{VV_{\phi\phi}}{V_\phi^2}\,.
\label{Gam}
\end{eqnarray}
{}From Eqs.~(\ref{Hubble}) and (\ref{Hubbleeq}) we have
\begin{eqnarray}
\label{dotH}
\frac{H'}{H}=-\frac32 \left[\gamma-
\frac{(\gamma-x^2)y^2}{\sqrt{1-x^2}}\right]\,.
\end{eqnarray}

We also define
\begin{eqnarray}
\Omega_\phi&& \equiv \frac{\rho_\phi}{\rho_{\rm cr}} =
\frac{V(\phi)}{3M_p^2H^2 \sqrt{1-\dot{\phi}^2}}
=\frac{y^2}{\sqrt{1-x^2}}\,, \\
& &\Omega_B \equiv \frac{\rho_B}{\rho_{\rm cr}} =
\frac{\rho_B}{3M_p^2H^2}\,,
\end{eqnarray}
where $\rho_{\rm cr}=3M_p^2H^2$.
Note that these satisfy the constraint equation
$\Omega_\phi+\Omega_B=1$
by Eq.~(\ref {Hubbleeq}).
Since $0 \le \Omega_\phi \le 1$, the allowed
range of $x$ and $y$ is $0 \le x^2+y^4 \le 1$.
Therefore both $x$ and $y$ are finite in the range
$0 \le x^2 \le 1$ and $0 \le y \le 1$.
The effective equation of state for
the field $\phi$ is
\begin{eqnarray}
\label{gamphi}
\gamma_\phi=\frac{\rho_\phi+p_\phi}
{\rho_\phi}=\dot{\phi}^2\,,
\end{eqnarray}
which means that $\gamma_\phi \ge 0$.
The condition for inflation corresponds to
$\dot{\phi}^2<2/3$ \cite{Tachyonindustry}.
It is also convenient to introduce a standard
deceleration parameter as
\begin{eqnarray}
\label{q}
q \equiv -\frac{\ddot{a}a}{\dot{a}^2}
=\frac32 \gamma -1 -\frac32
\frac{\gamma- x^2}{\sqrt{1-x^2}}y^2\,.
\end{eqnarray}

Equation (\ref{auto}) implies that
one can have $\lambda={\rm const}$
for $\Gamma=3/2$.
In this case integrating Eq.~(\ref{Gam}) gives
\cite{Paddy,Bagla}
\begin{eqnarray}
V(\phi) =M^2\phi^{-2}\,.
\label{inversesqu}
\end{eqnarray}
This corresponds to the potential for scaling solutions 
in the context of braneworld cosmology \cite{Maeda01,MM,SS}.
Recently a phase-space analysis was performed in 
Ref.~\cite{AL} for the same potential in the tachyon system.
In this paper, we are intend to broaden this type of investigation 
to allow for the more general situation 
where the potential is not restricted to Eq.~(\ref{inversesqu}).
In fact the tachyon potentials we introduce are motivated either by string theory or
based  purely on phenomenological considerations.

For example, in the case of a general inverse power-law 
potential \cite{abdalla,AF}
\begin{eqnarray}
V(\phi) =M^{4-n}\phi^{-n}\,,
\label{inverse}
\end{eqnarray}
one has $\lambda \propto \phi^{(n-2)/2}$ and
$\Gamma=(n+1)/n$.
Therefore $\lambda$ is constant for $n=2$,
but it dynamically changes for $n \ne 2$.
In the limit of $\phi \to \infty$
we have $\lambda \to 0$ for $0<n<2$ and
$\lambda \to \infty$ for $n>2$.

It is convenient to classify the potential
$V(\phi)$ depending on the asymptotic behavior of
$\lambda$ as in the case of a normal scalar field \cite{MP}. 
They can be classified as follows:

\begin{itemize}
\item  (i) $\lambda={\rm const}$.

In the tachyon system, the inverse square potential
$V(\phi)=M^2\phi^{-2}$ gives a constant $\lambda$.
Note that an exponential potential corresponds to
a constant value of $\tilde{\lambda} \equiv -M_pV_\phi/V$
in the case of a normal scalar field \cite{CLW}.

\item (ii) $\lambda \to 0$ asymptotically.

There exist a number of potentials that exhibit this behavior.
For example
\begin{itemize}
\item (iia) $V=M^{4-n}\phi^{-n}$ with $0<n<2$.

In this case $\lambda$ ($\propto \phi^{(n-2)/2}$) approaches
0 as $\phi \to \infty$. It is known that inflation occurs
for $0<n<2$ \cite{AF}.

\item (iib) $V=V_0e^{1/(\mu \phi)}$.

This is the potential that is used in the quintessence
scenario \cite{Paul}, and $\lambda \propto e^{-1/(2\mu \phi )}/\phi^2$,
which satisfies $\lambda \to 0$ as $\phi \to \infty$.

\item (iic) $V(\phi)=V_0e^{\frac12 M^2\phi^2}$.

This potential may appear as the excitation of massive scalar
fields on the D-brane \cite{GST} and it has a minimum at $\phi=0$.
Since $\lambda \propto \phi e^{-\frac14 M^2\phi^2}$ in this case,
$\lambda \to 0$ as $\phi \to 0$.

\end{itemize}

\item (iii) $|\lambda| \to \infty$ asymptotically.

There are also several potentials of interest that give this behavior

\begin{itemize}

\item (iiia) $V=M^{4-n}\phi^{-n}$ with $n>2$.

As seen earlier, in this case $\lambda$ ($\propto
\phi^{(n-2)/2}$) approaches $\infty$ as $\phi \to \infty$.

\item (iiib) $V=V_0e^{-\mu \phi}$.

This potential was considered in Ref.~\cite{Sami02} in the 
context of tachyon inflation. 
Unlike the case of conventional cosmologies, in 
tachyon cosmology, the exponential potential does not lead to a
constant $\lambda$. Since $\lambda$ is proportional to
$e^{\frac12 \mu \phi}$, we find  $\lambda \to \infty$ as $\phi
\to \infty$.
This potential may appear in the late time behaviour of D$_3$
anti-D$_3$ cosmology \cite{KK1}. The tachyon in the D$_3$
anti-D$_3$ system is a complex scalar field $T=\phi e^{i\theta}$
\cite{sen2}. When $\theta$ is constant, the effective
action of coincident D$_3$ anti-D$_3$ is the same as
Eq.~(\ref{action}) \cite{sen1}. The tachyon potential for the
coincident D$_3$ anti-D$_3$ is the same as the tachyon potential
of non-BPS D$_3$-brane which is \cite{Liu}
$V(\phi)=2\beta^2T_3/\cosh(\sqrt{\beta}m\phi)$ where $\beta$ is
a warp factor at the position of the D$_3$ anti-D$_3$ in the
internal compact space, $T_3$ is the tension of branes and $m$ is
the string mass scale. At the late time ($\phi \to \infty$) the
potential behaves as $V(\phi)\sim \beta^2 T_3
e^{-\sqrt{\beta}m\phi}$.

\item (iiic) $V(\phi)=V_0e^{-\frac12 M^2\phi^2}$.

This is the case in which the tachyon rolls down toward
$\phi \to \infty$ unlike the potential (iic).
$\lambda$ has a dependence $\lambda =M_p M^2 V_0^{-1/2}  
\phi e^{\frac14 M^2\phi^2}$
thereby giving $\lambda \to \infty$ as $\phi \to \infty$.

\end{itemize}

\vspace{0.2cm}
It may be noted that the potentials (iiib) $\&$ (iiic) listed 
above can be obtained in the frame work of string tachyons
whereas the inverse power-law type potentials (iiia) are motivated 
from purely phenomenological considerations.

\end{itemize}

We can now summarise the above behaviour in terms of conditions
on $V(\phi)$. The system approaches $\lambda \to 0$ when the
slope of the potential is less steep than that of
$V=M^2\phi^{-2}$ and the field $\phi$ evolves toward infinity
without oscillations [(iia) and (iib)]. 
The case (iic) corresponds to the one  in which the field $\phi$ 
oscillates as it evolves toward zero asymptotically.

On the other hand, we have an asymptotic value $|\lambda| \to
\infty$ when the potential is steeper than that of
$V=M^2\phi^{-2}$ and the field $\phi$ evolves toward infinity
without oscillations [(iiia), (iiib) and (iiic)]. One may
consider the potential $V(\phi)=V_0(\phi-\phi_0)^n$ , 
for $n$ positive, that has a dependence $\lambda \propto
(\phi-\phi_0)^{-(n/2+1)}$, thereby showing a divergence of
$\lambda$ for $\phi \to \phi_0$. However this potential has a
number of problems such as a divergent negative mass $m_{\rm
eff}^2=({\rm log}\,V)_{\phi \phi}$ as $\phi \to \phi_0$, which
leads to a violent instability of perturbations in the context of
tachyon cosmology \cite{Frolov}. We do not regard this as a
realistic dark energy tachyon potential. Note that if the
potential has a positive constant energy at $\phi=\phi_0$ then
this system reduces to that of the case (iic).

\section{Constant $\lambda$}

Let us first consider the situation in which $\lambda$
is constant, i.e., case (i) in the previous section.
The fixed points for this system can be obtained by setting $x'=0$
and  $y'=0$ in Eqs.\,(\ref{dotx}) and (\ref{doty}).
These are summarized in Table I.
Essentially we have 4 fixed points: (a) $x=0$, $y=0$,
(b) $x=\pm 1$, $y=0$, (c) $x=\lambda y_s/\sqrt{3}$,
$y=y_s$ and (d) $x=\sqrt{\gamma}$,
$y=\pm \sqrt{3\gamma}/\lambda$.
Here $y_s$ is defined by
\begin{eqnarray}
y_s=\left(\frac{\sqrt{\lambda^4+36}-\lambda^2}{6}
\right)^{1/2}\,.
\label{ys}
\end{eqnarray}
The cases (b) and (d) are divided into two cases, respectively,
depending on the signs of $x$.
The cases (c) and (d) correspond to stable fixed points.
These satisfy the conditions $3x=\sqrt{3}\lambda y$
and $3\gamma=\sqrt{3}
\lambda xy+\frac{3(\gamma -x^2)y^2}
{\sqrt{1-x^2}}$ in Eqs.~(\ref{dotx}) and  (\ref{doty}).

\begin{table*}[t]
\begin{center}
\begin{tabular}{|c|c|c|c|c|c|c|}
Name & $x$ & $y$ & Existence & Stability & $\Omega_\phi$
 & $\gamma_\phi$ \\
\hline
\hline
(a) & 0 & 0 & All $\lambda$ and $\gamma$ & Unstable saddle for
$\gamma > 0$  &   0 & 0 \\
& & & & Stable node for $\gamma=0$ & & \\
\hline
(b1) & 1 & 0 & All $\lambda$ and $\gamma$ & Unstable node
& 1 & 1 \\
\hline
(b2) & $-1$ & 0 & All $\lambda$ and $\gamma$ & Unstable node
& 1 & 1 \\
\hline
(c) & $\lambda y_s/\sqrt{3}$ & $y_s$ & All $\lambda$ and $\gamma$ &
Stable node for $\gamma \ge \gamma_s$ & 1 & $\lambda^2 y_s^2/3$ \\
 & & & & Unstable saddle for $\gamma < \gamma_s$ & & \\
\hline
(d1) & $\sqrt{\gamma}$ & $\sqrt{3\gamma}/\lambda$
& $\lambda>0$ and $\gamma<\gamma_s$ & Stable node & $\frac{3\gamma}{\lambda^2}
\frac{1}{\sqrt{1-\gamma}}$ & $\gamma$  \\
\hline
(d2) & $-\sqrt{\gamma}$ & $-\sqrt{3\gamma}/\lambda$
& $\lambda<0$ and $\gamma<\gamma_s$ & Stable node & $\frac{3\gamma}{\lambda^2}
\frac{1}{\sqrt{1-\gamma}}$ & $\gamma$
\end{tabular}
\end{center}
\caption[crit]{\label{crit} The critical points for constant $\lambda$}
\end{table*}

\subsection{Stability of the fixed point solutions}

We now study the stability around the critical points
given in Table I.
Consider small perturbations $u$ and $v$ about the
points $(x_c, y_c)$, i.e.,
\begin{eqnarray}
x=x_c+u\,,~~~y=y_c+v\,.
\label{uv}
\end{eqnarray}
Substituting into Eqs.~(\ref{dotx}) and (\ref{doty}),
leads to the first-order differential equations:
\begin{eqnarray}
\left(
\begin{array}{c}
u' \\
v'
\end{array}
\right) = {\cal M} \left(
\begin{array}{c}
u \\
v
\end{array}
\right) \ ,
\label{uvdif}
\end{eqnarray}
where ${\cal M}$ is a matrix that depends upon
$x_c$ and $y_c$.

The system can be regarded as being perturbatively stable when the
the eigenvalues of the matrix ${\cal M}$ are both
negative \cite{CLW}.
In what follows we shall obtain the eigenvalues $\mu_1$
and $\mu_2$ for
the fixed points in Table I and discuss their stability.

\begin{itemize}
\item
Case (a) [$x_c=0$, $y_c=0$]:

The eigenvalues are
\begin{eqnarray}
\mu_1=-3\,,~~~\mu_2=3\gamma/2\,.
\label{mua}
\end{eqnarray}
Therefore this critical point is an unstable saddle for
$\gamma>0$, whereas it is a stable node for $\gamma=0$.
This fixed point can not be used as a late-time attractor
solution, since it leads to  $\Omega_\phi=0$.

\item
Case (b) [$x_c= \pm 1$, $y_c=0$]:

Since the eigenvalues are
\begin{eqnarray}
\mu_1=6\,,~~~\mu_2=3\gamma/2\,,
\label{mub}
\end{eqnarray}
this fixed point is an unstable node.
This corresponds to a dust-like solution with 
$\gamma_\phi=\dot{\phi}^2=1$,
but the system tends to repel from this critical point.

\item
Case (c) [$x_c=\lambda y_s/\sqrt{3}$, $y_c=y_s$]:

The eigenvalues are
\begin{eqnarray}
\mu_1 &=&-3+\frac{\lambda^2}{12}
(\sqrt{\lambda^4+36}-\lambda^2)\,, \\
\mu_2 &=&-3\gamma+
\frac{\lambda^2}{6}
(\sqrt{\lambda^4+36}-\lambda^2)\,,
\label{muc}
\end{eqnarray}
where $\mu_1$ ranges between $-3 \le \mu_1 <-3/2$.
We have $\mu_ 2\le 0$ for
\begin{eqnarray}
\gamma \ge \gamma_s \equiv
\frac{\lambda^2}{18}
(\sqrt{\lambda^4+36}-\lambda^2)\,.
\label{gam}
\end{eqnarray}
This means that the fixed point is a stable node for
$\gamma \ge \gamma_s$, whereas  it is an unstable
saddle point for $\gamma<\gamma_s$.

\item
Case (d) [$x_c=\pm \sqrt{\gamma}$, $y_c=
\pm \sqrt{3\gamma}/\lambda$]:

The eigenvalues are
\begin{eqnarray}
\label{mu1mu2}
\mu_{1, 2}= \frac34 \biggl[\gamma-2
\pm \sqrt{17\gamma^2-20\gamma+4+
\frac{48}{\lambda^2}\gamma^2\sqrt{1-\gamma}}
\biggr]. \nonumber \\
\end{eqnarray}
The real parts of $\mu_1$ and $\mu_2$ are both
negative if the condition
\begin{eqnarray}
0 \le \gamma \le \gamma_s=\frac{\lambda^2}{18}
(\sqrt{\lambda^4+36}-\lambda^2)\,,
\end{eqnarray}
is satisfied. Note that $\gamma_s$ is always smaller than 1.
When the square root in Eq.~(\ref{mu1mu2}) is positive,
the fixed point is a stable node.
The fixed point is a stable spiral  when
the square root in Eq.~(\ref{mu1mu2}) is negative.

The values $\Omega_\phi$ and $\gamma_\phi$
at the critical point are
\begin{eqnarray}
\label{sca}
\Omega_\phi=\frac{3\gamma}{\lambda^2}
\frac{1}{\sqrt{1-\gamma}}\,,~~~
\gamma_\phi=\gamma\,,
\end{eqnarray}
which corresponds to a scaling solution in which the energy
densities $\rho_\phi$ and $\rho_B$ decrease with the same rate.
However we need to caution the reader in that the scaling solution
does not exist in the either the matter ($\gamma=1$)
or radiation dominated ($\gamma=4/3$) eras, because
the existence of the scaling solution requires the
condition $0 \le \gamma \le \gamma_s<1$.
In this sense this solution can not be applied as a realistic
model of dark energy.

\end{itemize}

\subsection{The existence of scaling solutions}

It was shown in Ref.\,\cite{PT} that the existence of
scaling solutions restricts the form of
the Lagrangian of a scalar field $\vp$ to be
\begin{eqnarray}
\label{pgene}
p(X, \vp)=X\,g(Xe^{\lambda \vp})\,,
\end{eqnarray}
where $X=-g^{\mu \nu}\partial_\mu \vp \partial_\nu
\vp/2$ and $g$ is any function of $Xe^{\lambda \vp}$.
Equation~(\ref{pgene}) was derived by starting from a
general Lagrangian $p(X, \vp)$ which is an arbitrary
function of $X$ and $\vp$.
Recently this result was extended to a more general background 
given by $H^2 \propto \rho^n$ \cite{SS}.

The Lagrangian of our tachyon system is
\begin{eqnarray}
\label{tachlag}
p(X, \phi)=-V(\phi)\sqrt{1-2\tilde{X}}\,,
\end{eqnarray}
where $\tilde{X}=-g^{\mu \nu}\partial_\mu \phi
\partial_\nu \phi/2$.
At first glance it seems that this lagrangian
does not satisfy the condition for the existence
of scaling solutions given in Eq.\,(\ref{pgene}).
However we have earlier shown that this tachyon system
(\ref{tachlag}) does in fact possess scaling solutions
given by Eq.~(\ref{sca}).

One can clarify the situation by expressing the Lagrangian
(\ref{pgene}) in terms of the variable $\phi
\equiv (2/\lambda)\,e^{\lambda \vp/2}$.
Then Eq.~(\ref{pgene}) becomes
\begin{eqnarray}
\label{lagnew}
p(\tilde{X}, \phi)=\frac{4}{\lambda^2 \phi^2}
f(\tilde{X})\,,
\end{eqnarray}
where $f(\tilde{X}) \equiv \tilde{X}\,g(\tilde{X})$.
The tachyon system (\ref{tachlag})
can be accommodated by choosing
\begin{eqnarray}
V(\phi) \propto \phi^{-2}\,,~~~
f(\tilde{X}) \propto \sqrt{1-2\tilde{X}}\,.
\end{eqnarray}
Therefore scaling solutions exist for the Lagrangian
(\ref{tachlag}) in the case of the inverse square
potential, as of course we already knew.
It is interesting to note that
Eq.~(\ref{pgene}) completely fixes
the form of the Lagrangian for the existence of
scaling solutions \cite{PT,SS}.

\subsection{Late time behavior}

The cosmological dynamics of the tachyon field with inverse square potential
(\ref{inversesqu})
was studied in Ref.~\cite{Bagla,AL}. 
In what follows, we would like to clarify several important points
concerning the  application of this model to dark energy.

Employing slow-roll approximations:
$3H\dot{\phi} \simeq -V_\phi/V$ and $3M_p^2H^2 \simeq V(\phi)$
in a scalar-field dominated universe ($\rho_\phi \gg \rho_B$),
we obtain
\begin{eqnarray}
\label{phisolu}
& &\dot{\phi}=\frac{2M_p}{\sqrt{3}M}\,, \\
& & a \propto t^p\,,~~{\rm with}~~
p \equiv \frac12 \left(\frac{M}{M_p}
\right)^2\,,
\end{eqnarray}
for the potential (\ref{inversesqu}).
In order to have acceleration at late times, we clearly require
$p>1$, i.e.,
\begin{eqnarray}
M>\sqrt{2}M_p\,.
\end{eqnarray}
Furthermore $M$ needs to be much larger than the Planck
mass to obtain a significant acceleration ($p \gg 1$).
Such a large mass is problematic as we expect general relativity itself to 
break down in such a regime. This problem is fortunately alleviated
 for the inverse power-law potential
$V=M^{4-n}\phi^{-n}$ with $0<n<2$, as we will see later.

When $n=2$ the parameter $\lambda$ defined in
Eq.~(\ref{lam}) is a constant and given by
\begin{eqnarray}
\lambda=2\frac{M_p}{M}=\sqrt{\frac{2}{p}}\,.
\end{eqnarray}
Then we require $\lambda<\sqrt{2}$ for the acceleration.
The fixed point (c) in Table I is the only stable attractor
solution with an accelerating universe at late times
(recall that the fixed point (d) is not a realistic solution).
Let us consider the case of $\lambda \ll 1$ (i.e., $p \gg 1$).
The fixed point (c) is approximately given as
\begin{eqnarray}
\label{xycri}
x \simeq \frac{\lambda}{\sqrt{3}}\,,~~~
y \simeq 1-\frac{\lambda^2}{12}\,,
\end{eqnarray}
together with the equation of state
\begin{eqnarray}
\gamma_\phi  \simeq \frac{\lambda^2}{3}=\frac{2}{3p}\,.
\end{eqnarray}
The condition for acceleration ($p>1$) translates
into $\gamma_\phi<2/3$.
One can easily verify that the slow-roll solution
in Eq.~(\ref{phisolu}) is identical to the critical
point (c) given in Eq.~(\ref{xycri}).
Therefore the slow-roll solution (\ref{phisolu})
is a stable attractor which gives $\Omega_\phi=1$
and $\gamma_\phi=2/(3p)$.

We note that there is another constraint coming from
the energy scale of the tachyon potential.
Since the energy density of the tachyon is supposed to
be the same order as the present critical density $\rho_{\rm cr}
=10^{-47}\,{\rm GeV}^4$, we require the condition
$M^2\phi_0^{-2} \simeq \rho_{\rm cr}$.
This restricts the present field values to be
$\phi_0 M \gtrsim 10^{60}$.

\section{Case of $\lambda \to 0$}

In this section we shall study the case in which
$\lambda$ dynamically approaches 0.
Examples of  potentials which exhibit this
behavior was presented in section\, II.
In the limit of $\lambda \to 0$ Eqs.~(\ref{dotx})
and (\ref{doty}) read
\begin{eqnarray}
\label{dxlam0}
& &x'=-3x(1-x^2)\,, \\
& &y'=-\frac{H'}{H}y\,.
\label{dylam0}
\end{eqnarray}
By Eq.~(\ref{dxlam0}) $x'<0$ for $x>0$ and
$x'>0$ for $x<0$ (note that $x^2$ is in the range
$0 \le x^2 \le 1$).
Therefore $x \to 0$.
Integrating Eq.~(\ref{dylam0}) gives $y \propto 1/H$,
which means that $y$ continues to increase
toward $y=1$ as $H$ decreases.
Then the attractor solution should correspond to
$x=0$ and $y=1$.
This discussion neglects the contribution from terms like
$\sqrt{3}\lambda y$, but the solution actually approaches
the attractor $(x, y)=(0, 1)$ for $\lambda \to 0$
as we see below.

While the fixed points we found in the previous
section correspond to the case of constant $\lambda$,
we may regard these as ``instantaneous'' critical points
as the function $\lambda(N)$ evolves
toward 0.
The only stable critical point leading to an acceleration
at late times is that of case (c) in Table I, where
the dynamical critical point around $\lambda(N)=0$
is approximately given by Eq.~(\ref{xycri}).
This approaches $x \to 0$ and $y \to 1$ as $\lambda \to 0$.
Note also that $\gamma_\phi \to 0$
as  $\lambda \to 0$, which corresponds to
the equation of state of a cosmological constant.
We would like to numerically
confirm that the system actually approaches
the dynamical critical point.
In what follows we classify the situation in two classes:
(A) $\lambda \to 0$ without any oscillations of $\phi$ and
(B) $\lambda \to 0$ with  $\phi$ oscillating.

\subsection{$\lambda \to 0$ with no oscillations of $\phi$}

The potentials which lead to this behaviour correspond to
(iia) $V(\phi)=M^{4-n}\phi^{-n}$ with $0<n<2$ and
(iib) $V(\phi)=V_0e^{1/(\mu \phi)}$. In this class of
models the field $\phi \to \infty$
without oscillations. As long as the tachyon potential is not
steep relative to the inverse square potential, i.e., $n<2$,
$\lambda(N)$ asymptotically approaches 0 as
$\phi$ increases. This means that the equation of state of the
field $\phi$ approaches $\gamma_\phi \simeq
\lambda(N)^2/3 \to 0$, which leads to the universe accelerating at late times.

Let us consider the inverse power-law potential
$V(\phi)=M^{4-n}\phi^{-n}$.
In this case one has $\Gamma=(n+1)/n$ in Eq.~(\ref{Gam}),
which means that $\lambda$ continues to decrease
for $0<n<2$.
The slow-roll parameter for the tachyon-type scalar
field is \cite{Tachyonindustry}
\begin{eqnarray}
\label{slowpara}
\epsilon =\frac{M_p^2}{2} \left(\frac{V_\phi}{V}
\right)^2 \frac{1}{V}
=\frac{n^2}{2} \left(\frac{M_p}{M}\right)^2
\frac{1}{(\phi M)^{2-n}}\,.
\end{eqnarray}
When $0<n<2$, $\epsilon$ decreases as the field evolves
toward large values.
The condition for the accelerated expansion corresponds
to $\epsilon<1$, which yields
\begin{eqnarray}
\label{phiM}
\phi M>\left(\frac{n}{\sqrt{2}} \frac{M_p}{M}
\right)^{2/(2-n)}\,.
\end{eqnarray}
If this tachyon field is to be responsible for the observed inflation today, then it must satisfy Eq.~(\ref{phiM}).

The present potential energy is approximated as
$V(\phi_0)=M^4/(\phi_0 M)^n \simeq \rho_{\rm cr} \simeq
10^{-47} {\rm GeV}^4$.
Combining this relation with Eq.~(\ref{phiM}) we get
\begin{eqnarray}
\label{Mcons}
\frac{M}{M_p}>\left[\left(\frac{\rho_c}{M_p^4}\right)
^{1-n/2}\left(\frac{n}{\sqrt{2}}\right)^n\right]^{1/(4-n)}\,.
\end {eqnarray}
The r.h.s. becomes smaller as $n$ decreases.
For example one has $M/M_p \gtrsim 10^{-20}$ for $n=1$.
Therefore the super-Planckian problem for the
inverse square potential is alleviated for $n<2$.

\begin{figure}
\includegraphics[height=8.3in,width=3.2in]{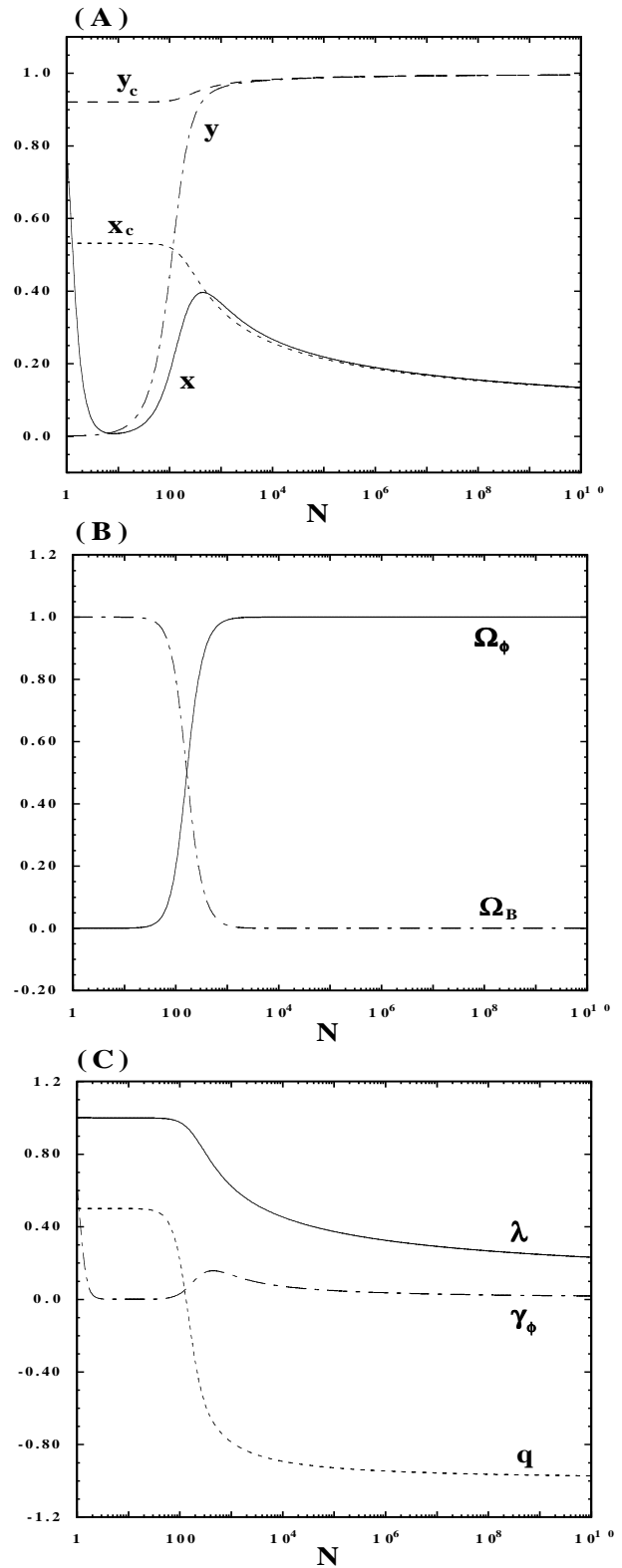}
\caption{\label{figure1}
Cosmological dynamics for the field $\phi$ with potential
$V(\phi)=M^3\phi^{-1}$ and a barotropic fluid
with $\gamma=1$.
We choose initial conditions as
$x_i=0.8$, $y_i=5.0 \times 10^{-4}$
and $\lambda_i=1.0$.
Each panel corresponds to the evolution
with (A) $x$, $y$ and the critical points $x_c$, $y_c$
for the case (c) in Table I, (B) $\Omega_\phi$, $\Omega_B$
and (C) $\lambda$, $\gamma_\phi$ and the
deceleration parameter $q$.
}
\end{figure}

In Fig.~\ref{figure1} we plot the cosmological evolution for
$n=1$ with initial conditions $x_i=0.8$,
$y_i=5.0 \times 10^{-4}$ and $\lambda_i=1.0$.
We consider a pressureless dust ($\gamma=1$)
as a background fluid.
Note that $\epsilon$ and $\lambda$ satisfy the relation
$\epsilon=\lambda^2/2$ by Eqs.~(\ref{lam}) and
(\ref{slowpara}). Our choice $\lambda_i=1$
corresponds to the initial condition $\epsilon_i=0.5$.
This does not mean that inflation occurs at the initial
stage, since the energy density of the barotropic fluid
dominates over that of the scalar field.
When the energy density of $\phi$ eventually wins out, this leads
to the acceleration of the universe since the slow-roll
parameter is smaller than of order 1.

{}From the panel (A) of Fig.\,\ref{figure1} we find that $x$ decreases initially.
This comes from the fact that the condition $3x \gg
\sqrt{3}\lambda y$ holds in Eq.~(\ref{dotx}) for our initial
conditions. On the other hand $y$ grows through the relation
$y' \approx (3\gamma/2)y$. Since this growth is rather rapid
and $\lambda$ is nearly constant initially,
the $\sqrt{3}\lambda y$ term temporarily surpasses
the $3x$ term in Eq.~(\ref{dotx}), which leads to the
increase of $x$ for a short period.
After that, $\lambda$ begins to decrease [see the panel (C)]
and this has the effect of balancing the two terms
($3x \approx \sqrt{3}\lambda y$).
In Fig.\,\ref{figure1} we plot the dynamically changing critical points
(c) in Table I, i.e., $(x_c, y_c)=(\lambda(N) y_s(N)/\sqrt{3}, y_s(N))$
with $y_s(N)=[(\sqrt{\lambda(N)^4+36}-\lambda(N)^2)/6]^{1/2}$.
We find that the solution approaches these ``instantaneous'' critical
points with the decrease of $x$.
Therefore the discussion of constant $\lambda$ can be applied to the case
of varying $\lambda$ after the system approaches the stable attractor
solutions.
One can find the asymptotic evolution of $\lambda$ by substituting
Eq.~(\ref{xycri}) for Eq.~(\ref{auto}). This gives the dependences
$\lambda \propto 1/\sqrt{N}$, $x \propto 1/\sqrt{N}$ and
$1-y \propto 1/N^2$, which we confirmed numerically.
Note that this relation holds as long as $\Gamma$ is asymptotically
constant with $\Gamma<3/2$.

The evolution of $\Omega_\phi$, $\Omega_B$, $\lambda$, $\gamma_\phi$
and $q$ is plotted as well in the panels (B) and (C) in Fig.~\ref{figure1}.
The deceleration parameter $q$ becomes negative for $N \gtrsim 100$,
after which the system enters the acceleration stage
with the growth of $\Omega_\phi$.
We checked that $\gamma_\phi$ evolves toward 0 with the decrease of
$\lambda(N)$ keeping the relation $\gamma_\phi \simeq \lambda(N)^2/3$.

We can also account for a combined system of two fluids (matter and
radiation) with a scalar field $\phi$. We have done this, running
our numerical code from
a redshift $z=10^6$, finding that it is possible to obtain a viable
cosmological evolution for the inverse power-law potential with $n<2$.
The basic property of the dynamical system for the potential
$V(\phi)=V_0e^{1/(\mu \phi)}$ is similar to what we discussed
above.

\subsection{$\lambda \to 0$ with oscillations of $\phi$}

One example exhibiting this  type of behaviour is the case (iic),
i.e., a rolling massive scalar field with potential $V(\phi)=V_0
e^{\frac12 M^2\phi^2}$ \cite{GST}. This potential has a minimum
at $\phi=0$ with an energy density $V_0$. As an example, consider
an anti-D3 brane at the tip of Klebanov-Strassler (KS) throat 
in the KKLT setup \cite{KKLT}. Since
there is a warp factor $\beta$ at this point, the potential of a
massive excitation of the anti-D$_3$-brane $\phi$ may be
written as $V(\phi)=\beta^2 T_3 e^{\frac12 \beta m^2\phi^2}$,
where $T_3$ is the brane tension and $m$ is the mass of the
excited state of the brane which is of order the string mass scale.
Hence, if we introduce a very small warp factor $\beta$, it
should be possible to explain the origin of the present dark
energy. Note that with a warp factor $\beta$ of order 1, the massive
scalar field decays to $\phi=0$ very soon in the reheating epoch.
However, for a very small warp factor, the stabilization of the field
$\phi$ is considerably delayed allowing it to play the role of dark
energy today.

Another possibility is to consider an anti-D$_3$-brane with a warp
factor $\beta'$ and a negative cosmological constant ($-\Lambda$)
arising from the stabilization of modulus fields in the KKLT
vacua with the assumption that the potential energy of the rolling
massive scalar does not exactly cancel the cosmological constant
($V(\phi)-\Lambda \simeq \rho_{\rm cr}$) \cite{GST}. The rolling  
massive scalar, in order to play the roll of quintessence, is expected to
be stabilized in future.

\begin{figure}
\includegraphics[height=8.3in,width=3.2in]{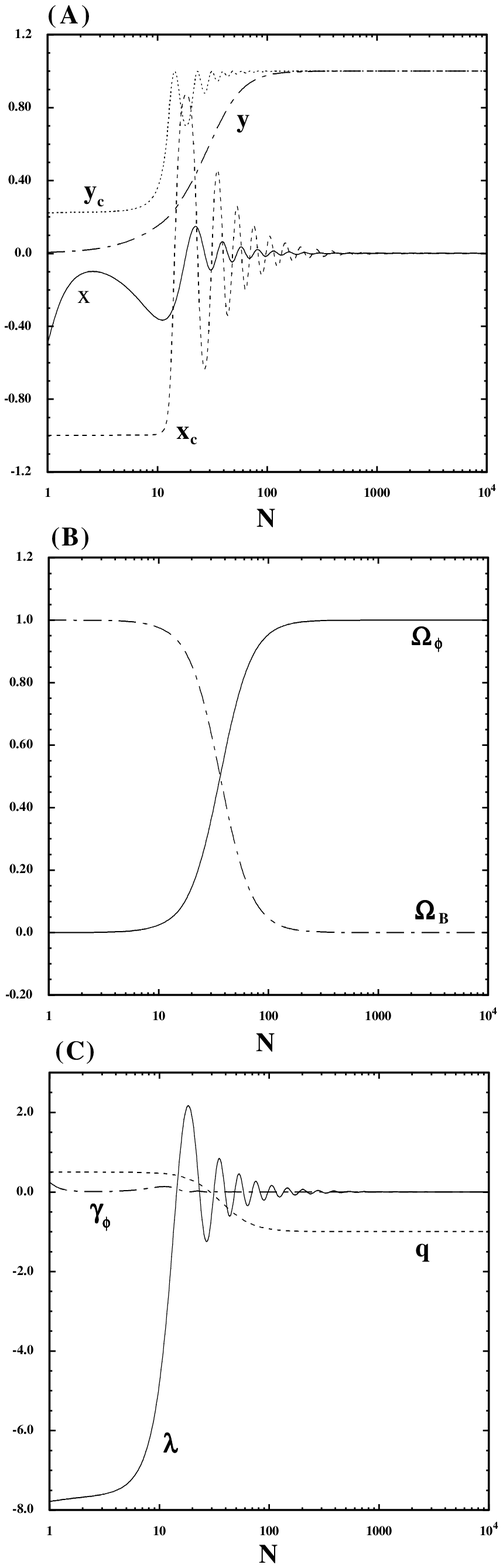}
\caption{\label{figure2}
Cosmological dynamics for the field $\phi$ with
potential $V(\phi)=V_0e^{\frac12 M^2 \phi^2}$
and a barotropic fluid with $\gamma=1$.
The initial conditions are $x_i=-0.5$,
$y_i=6.7 \times 10^{-3}$ and $\lambda_i=-7.8$.
We show the evolution of the same quantities as in Fig.\,1.
}
\end{figure}

In this work we shall simply adopt the potential
$V(\phi)=V_0 e^{\frac12 M^2\phi^2}$
and study the dynamics of the system when the field
oscillates around the potential minimum.
Since $\lambda=-M_pM^2\phi/
\sqrt{V_0}e^{-\frac14M^2\phi^2}$, this quantity gradually
decreases toward 0 and is accompanied with the oscillations of $\phi$.
In Fig.\,\ref{figure2} we plot one example of the
cosmological evolution for this
scalar potential.
The quantity $x$ approaches 0 with damped oscillations
as expected. We find that the ``instantaneous'' critical
points (c) in Table I provide a good description of the late time evolution of $x$ and $y$.
$\Omega_\phi$ begins to grow toward 1 around
$N \approx 20$, after which the system enters the
accelerating stage [see the evolution of $q$ in the panel (C)].

\section{Case of $|\lambda| \to \infty$}

There are a number of potentials that give $|\lambda| \to \infty$
asymptotically. One can classify the tachyon potentials into two
classes: (1) $|\lambda| \to \infty$ without the oscillation of
the field $\phi$ [(iiia), (iiib), (iiic)] and (2) $|\lambda| \to
\infty$ with the oscillation of the field, that is,
$V(\phi)=V_0(\phi-\phi_0)^n$ for positive $n$. In the latter case
the effective mass squared $m_{\rm eff}^2=({\rm log}\,V)_{\phi
\phi}$ of the field $\phi$ is negative, which leads to a violent
instability for the tachyon perturbations \cite{Frolov}. This is
not regarded as a stable attractor unlike the potential
$V(\phi)=V_0e^{\frac12 M^2\phi^2}$ discussed in the previous
section.

\begin{figure}
\includegraphics[height=8.3in,width=3.2in]{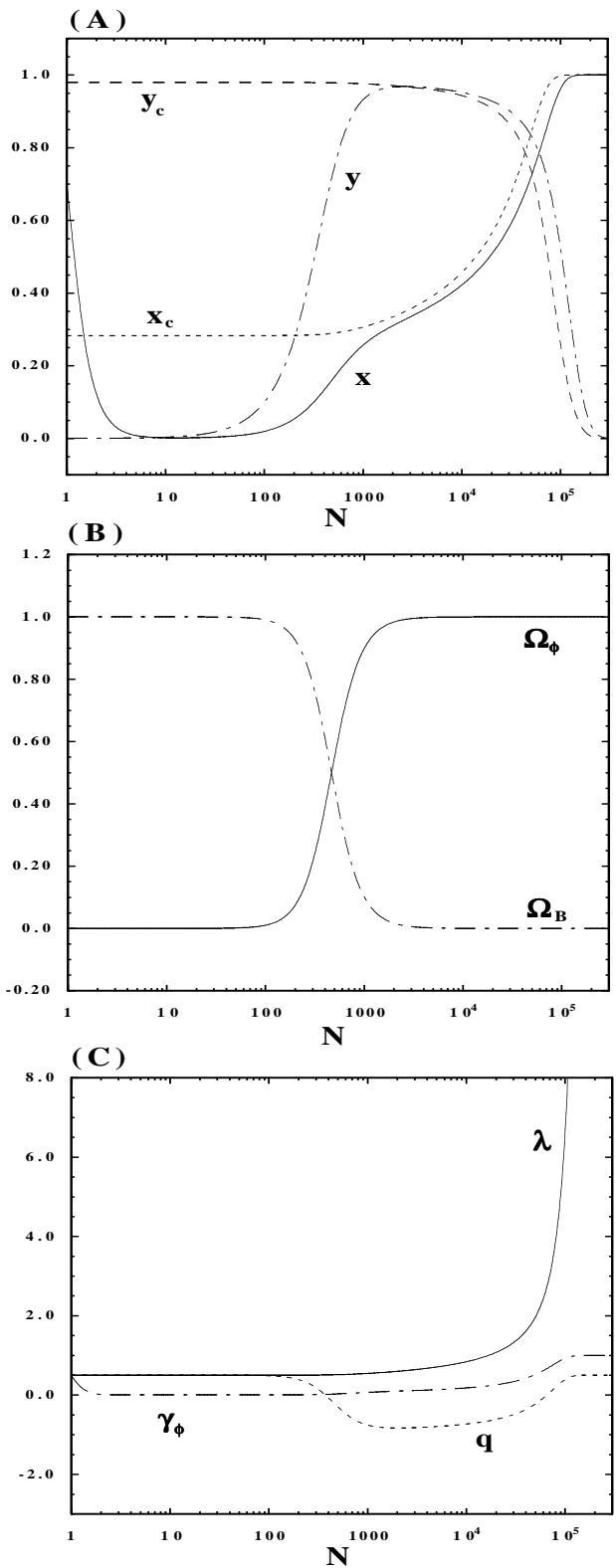}
\caption{\label{figure3}
Cosmological evolution for the field $\phi$ with
potential $V(\phi)=V_0e^{-\mu \phi}$
and the barotropic fluid with $\gamma=1$.
Initial conditions are $x_i=0.7$, $y_i=1.0 \times 10^{-4}$
and $\lambda_i=0.5$.
We show the evolution of the same quantities as in Fig.\,1.
}
\end{figure}

We shall investigate the case (1) in which $|\lambda|$
continues to grow asymptotically.
For example let us consider the exponential potential
$V(\phi)=V_0e^{-\mu \phi}$ with $\mu>0$ [case (iiib)].
Since $\Gamma=1$ for this potential, we have
$\lambda'=(\sqrt{3}/2)\lambda^2xy$, thereby leading
to the growth of $\lambda$ for $x>0$.
In Fig.\,\ref{figure3} we plot the cosmological evolution
for this system
with initial conditions $x_i=0.7$, $y_i=1.0 \times 10^{-4}$
and $\lambda_i=0.5$. There is a short initial stage in which
the conditions, $x \ll 1$ and $y \ll 1$, are satisfied with
$\Omega_\phi \simeq 0$ and $\gamma_\phi \simeq 0$.
This is an unstable fixed point (a) in Table I, thus
showing a deviation from $x=0$ and $y=0$ for
$N \gtrsim 100$. {}From the panel (C) we find that
the universe begins to accelerate for $N \gtrsim 400$
(see the evolution of $q$).
After that $x$ and $y$ approach the instantaneous critical
points (c) in Table I.
As long as $\lambda \leq {\cal O}(1)$, one can
see that the condition $q<0$ is satisfied by
substituting $x_s=\lambda y_s/\sqrt{3}$ and $y=y_s$
for Eq.~(\ref{q}).
With the growth of $\lambda$, however, the acceleration
of the universe stops for $N \gtrsim 6.5 \times 10^4$.
At this stage $x$ grows toward 1, where as $y$ decreases
toward 0. The asymptotic behavior corresponds to
$x \to 1$ and $y \to \sqrt{3}/\lambda$,
which can be obtained by taking the
limit $\lambda \gg 1$ for the instantaneous critical
points (c) in Table I. Substituting this solution for
Eq.~(\ref{auto}), we obtain
$\lambda \propto e^{3N/2}$. This relation holds
when $\Gamma$ is asymptotically constant
with $\Gamma>3/2$.
In this regime the equation of state for the field $\phi$
is characterized by $\gamma_\phi \simeq 1$ (i.e.,
a pressureless dust) with a dominant energy
density ($\Omega_\phi \simeq 1$).

It is worth reflecting here briefly on the fact that there exists 
a period of transient acceleration when we have 
an exponential potential.
Because of a dynamical change of $\lambda$,
it is possible to have a temporal acceleration for
$\lambda \lesssim 1$ and have a deceleration
for $\lambda \gg 1$.
If this temporal acceleration corresponds to the one
at present, the universe will eventually enter the
non-accelerating regime in which the tachyon
field behaves as a pressureless dust.  
This is a nice feature of the model 
which makes it free from the future event horizon problem
present in most of the quintessence models. 

As long as $|\lambda|$ approaches infinity asymptotically,
we do not have an acceleration of the universe
at late times.
This comes from the fact that the fixed point
(c) in Table I approaches $x \to 1$ and $y \simeq
\sqrt{3}/\lambda \to 0$ as $|\lambda| \to \infty$.
Therefore any potentials which exhibit this behavior
can not be used as a late-time dark energy candidate.
In order to lead to acceleration at late times, we
require that the tachyon potential has the property (i)
$\lambda={\rm const}$ or (ii) $\lambda \to 0$
asymptotically, as we showed in previous sections. 
Of course, as we have just mentioned, it is possible that 
the acceleration we are experiencing is not a late time evolution, 
but part of a transient regime!

\section{Summary}

In this paper we have studied the cosmology associated with
the tachyon field $\phi$ as the dark energy.
In the tachyon system the inverse square potential
(\ref{inversesqu}) is the marginal case which leads to an accelerating universe.
However, it is not the only possible tachyon potential and 
so we have deliberately set out to develop a unified analysis 
that can be applied for any type of tachyon potentials.

A crucial quantity to determine the cosmological dynamics
is $\lambda$ defined in Eq.~(\ref{lam}).
The inverse square potential (\ref{inversesqu})
corresponds to a constant $\lambda$
with $\lambda=2M_p/M$.
In this case we have essentially four critical points
as shown in Table I.
Among them a viable fixed point that gives a stable
attractor for dark energy is the case (c) with a barotropic
equation of state $\gamma$ larger than $\gamma_s$
[$\gamma_s$ is defined in Eq.~(\ref{gam})].
The critical point (d) corresponds to a scaling solution
in which the energy density of $\phi$ decreases
similarly to that of the barotropic fluid.
However this is not a viable cosmological scaling
solution, since the existence of it requires
the condition $\gamma<1$.

The tachyon potentials can be placed into three classes:
(i) $\lambda={\rm const}$, (ii) $\lambda \to 0$
asymptotically and (iii) $|\lambda| \to \infty$
asymptotically. The class (ii) corresponds to
the case in which the potential is not steep
relative to $V(\phi)=M^2\phi^{-2}$,
whereas the potential in the class (iii)
is steeper than $V(\phi)=M^2\phi^{-2}$.
While the former leads to the acceleration
at late times, the latter does not.

We have carried out a detailed analysis about the cosmological
evolution of the tachyon system in two cases: (1) $\lambda \to 0$
without the oscillation of $\phi$ and (2) $\lambda \to 0$
with the oscillation of $\phi$.
We adopt the inverse power-law potential
$V(\phi)=M^{4-n}\phi^{-n}$ with $0<n<2$
as an example of the case (1).
This is favorable relative to the inverse quadratic potential
($n=2$), since the mass scale $M$ is not severely constrained.
We solved the dynamical equations (\ref{dotx})-(\ref{auto})
numerically and found that the solutions approach
the ``instantaneous'' critical point (c) in Table I
with the decrease of $\lambda$ toward 0 [see
the panel (A) of Fig.\,1].
Since $x \simeq \lambda(N)/\sqrt{3} \to 0$ and
$y \simeq 1-\lambda(N)^2/12 \to 1$ as $\lambda(N) \to 0$,
the universe exhibits an acceleration at late times.
This is a dark energy scenario in which the future
universe is dominated by the field $\phi$ ($\Omega_\phi
\simeq 1$) with an equation of state $\gamma_\phi \simeq 0$.
We adopt the rolling massive scalar potential
$V(\phi)=V_0e^{\frac12 M^2\phi^2}$ to study the case (2)
mentioned above. We find that $x$ and $y$ again approach
the instantaneous critical point (c) with oscillations.
Since the potential has an energy $V_0$
at the potential minimum, this eventually
leads to an acceleration of the universe even if the
scalar field oscillates.

On the other hand we do not find an accelerated expansion
at late times when $|\lambda|$ grows towards infinity,
since the dynamical critical point (c) approaches
$x \to 1$ and $y \to 0$ as $|\lambda| \to \infty$.
Nevertheless we found that a transient acceleration can
occur in the region where $\lambda(N)$ is smaller
than of order unity (see Fig.\,3). It is tempting to speculate that 
this could be the transient regime we are experiencing today, 
and to try and predict when we expect it to change over again 
to a matter dominated era. Recall that the supernova data 
is not actually informing us about the evolution today ($z=0$), 
rather it tells us that the universe is accelerating at a redshift of
 order 0.05. Perhaps we are actually back in a matter dominated 
 regime today and just don't know it yet!

Inspite of the existing features of cosmological dynamics based upon DBI
scalars, it may happen that these models lead to the
formation of caustics where the second and the higher-order 
derivatives of the field become singular. 
We do not know whether caustics are generic prediction of string theory 
or appear as a result of the derivative truncation leading 
to the DBI action. As demonstrated in Ref.~\cite{FKS},
caustics inevitably form in tachyon system with potentials
decaying as $\phi^{-2}$ or faster at infinity. 
It remains to extend the analysis of Ref.~\cite{FKS} to the case 
of potentials such as $V \sim \phi^{-n}$ with $0<n<2$ 
analysed in our paper. Caustics normally form in systems with 
pressureless dust which is mimicked by tachyon field 
with run away potentials. It is therefore quite
likely that caustics may not develop in Born-Infeld systems 
with a ground state at at a finite value of the field. 
The rolling massive scalar potential 
$V(\phi)=V_0 e^{\frac12 M^2\phi^2}$ belongs
to this category. In our opinion, these are important issues which require
further investigations.

Finally we believe it is quite intriguing that the tachyon system 
provides rich and fruitful cosmological scenarios for dark
energy. The classification we have performed in this paper
provides a very useful way to find out about the cosmological evolution
for any type of tachyon potentials.

\section*{ACKNOWLEDGEMENTS}

We thank Shuntaro Mizuno for useful discussions.
E.C. is grateful to the Aspen Center for Physics for their hospitality 
during a period when some of this work was performed.
S.T. is grateful to Universities of Sussex, Queen Mary and
Portsmouth for their warm hospitality during which a
part of this work was done. 




\end{document}